# Decoherence window and electron-nuclear cross-relaxation in the molecular magnet $V_{15}$

J.H. Shim,[1*] S. Bertaina,[2] S. Gambarelli,[1] T. Mitra,[3§] A. Müller,[3] E.I. Baibekov,[4] B.Z. Malkin,[4] B. Tsukerblat,[5] and B. Barbara[1,6]

[1]Laboratoire de Chimie Inorganique et Biologique (UMR-E3 CEA-UJF), INAC, CEA-Grenoble, 17 Ave. des Martyrs, 38054 Grenoble Cedex 9, France

[2]IM2NP, CNRS, UMR 7334 & Université Aix-Marseille, Ave. Escadrille Normandie Niemen - Case 142 - 13397 Marseille Cedex 20, France

[3]Fakultät für Chemie, Universität Bielefeld, Postfach 100131, D-33501 Bielefeld, Germany

[4] Kazan Federal University, Kazan, 420008, Russian Federation

[5]Ben-Gurion University of the Negev, 84105 P.O. Box 653, Beer-Sheva, Israel

[6]Institut Néel, CNRS and Université Joseph Fourier, 25 Ave. des Martyrs, BP166, 38042 Grenoble Cedex 9, France

**Abstract**

Rabi oscillations in the $V_{15}$ Single Molecule Magnet (SMM) embedded in the surfactant DODA have been studied at different microwave powers. An intense damping peak is observed when the Rabi frequency $\Omega_R$ falls in the vicinity of the Larmor frequency of protons $\omega_N$, while the damping time $\tau_R$ of oscillations reaches values 10 times shorter than the phase coherence time $\tau_2$ measured at the same temperature. The experiments are interpreted by the *N*-spin model showing that $\tau_R$ is directly associated with the decoherence via electronic/nuclear spin cross-relaxation in the rotating reference frame. It is shown that this decoherence is accompanied with energy dissipation in the range of the Rabi frequencies $\omega_N - \sigma_e < \Omega_R < \omega_N$, where $\sigma_e$ is the mean super-hyperfine field (in frequency units) induced by protons at SMMs. Weaker damping without dissipation takes place outside this dissipation window. Simple local field estimations suggest that this rapid cross-relaxation in resonant microwave field observed for the first time in SMM $V_{15}$ should take place in other SMMs like $Fe_8$ and $Mn_{12}$ containing protons, too.

Compounds containing single Molecule Magnets (SMMs) are composed of a lattice of molecules or clusters whose collective ground-state spin $S$ is associated with strong intra-molecular interactions (essentially ferromagnetic as in $Mn_{12}$, $Fe_8$ with $S=10$ or antiferromagnetic as in $V_{15}$ with $S=1/2$). SMMs are considered as promising systems for quantum information processing because they can be self-organized in $2D$ or $3D$ networks, while their relatively small size (~1 nm) makes them good quantum objects with weak self-decoherence (which is generally non-controlled as in e.g. superconducting qubits). On the basis of both experiments and theory one can predict phase coherence times $\tau_2 = 100 \div 500$ μs in the SMMs $V_{15}$ [1] and $Fe_8$ [2] provided that nuclear spin-bath is absent and temperature and magnetic field are optimized. However, this takes place only in the absence of microwaves or when their application is confined to a short pulse sequence (as in $\tau_2$ measurements by means of the Hahn spin-echo method). Continuous application of resonant microwave field produces nutations of the spin magnetization (Rabi oscillations [3]). The relaxation dynamics of the spin system in this transient regime changes drastically. For a dipolar-coupled system of electronic spins, the Rabi time $\tau_R$ can become several times shorter than $\tau_2$ [4-6]. In-depth studies of decoherence in the presence of a sequence of short microwave pulses (as in a quantum calculation sequence) should allow testing these predictions. As opposed to common knowledge [7], under condition $\tau_R < \tau_2$, the utmost time of coherent spin manipulations in quantum computation will be limited by $\tau_R$ rather than $\tau_2$, and oscillations disappear if $\tau_R$ is less than the oscillation period.

First evidence of Rabi oscillations in a SMM was obtained in the anionic cluster $\left[V_{15}^{IV}As_6^{III}O_{42}(H_2O)\right]^{6-}$ (hereafter referred as $V_{15}$) embedded in the cationic surfactant DODA $\left[Me_2N\{(CH_2)_{17}Me_2\}\right]^+$ in order to reduce dipolar interactions [1, 8]. A follow-up study detected Rabi oscillations in another SMM, $Fe_4$ system [9]. In both cases several continuous oscillations were observed and the coherence times $\tau_2$ reached values of several hundred nanoseconds. Similar $\tau_2$ values were obtained in $Cr_7$-based [10] and more recently in $Fe_8$ [2] SMMs, but in these cases Rabi oscillations could not be detected. In this Letter we show both experimentally and theoretically that Rabi oscillations of $V_{15}$ – DODA are subjected to a very efficient and apparently general decoherence mechanism accompanied by energy dissipation into the proton spin bath. In a certain range of Rabi frequencies where this new phenomenon takes place, the condition $\tau_R < 2\pi/\Omega_R$ is nearly fulfilled, and Rabi oscillations almost disappear.

Each $V_{15}$ cluster has a layer structure with three $V^{IV}$ ions forming a large central triangle that is sandwiched by two smaller $V^{IV}_6$ hexagons [11, 12] (Fig. 4 in [12]). The 15 spins ½ of the $V^{IV}$ ions are coupled by strong antiferromagnetic super-exchange interactions in the external hexagons and by a relatively weak exchange through the bridges in the central triangle [11, 13, 14]. The lowest states of the energy spectrum are formed by a pair of doublets (with a gap of $\Delta_{1/2}$ = 200 mK caused by the Dzyaloshinskii-Moriya interactions [11, 14]) and by a quartet (with small zero-field splitting $\Delta_{3/2}$ = 12 mK [1, 11]). Exact diagonalization of the cluster Hamiltonian shows that these states are isolated from the above quasi-continuous spectrum by a gap of $\Delta$ = 250 K [11].

A series of $V_{15}$-DODA samples with concentrations $c = 1 \div 5\ c_0$ in units of $c_0 = 4.3 \cdot 10^{17}$ cm$^{-3}$ [15] was prepared following the method published by us earlier [1]. Variations of both the phase and spin-lattice relaxation times were measured *vs* cluster concentration (Fig. 6 in [12]). The observed linear slope of $d\tau_2/dc$ is apparently related to the decoherence via $V_{15}$ inter-cluster dipolar interactions. The phase coherence time measured at 4.2 K is of 0.5 μs for $c = c_0$ and decreases down to 0.25 μs for $c=4c_0$. The spin-lattice relaxation can be neglected at temperatures below 10 K where the spin-lattice relaxation time $\tau_1 > 10$ μs [12].

Rabi oscillations of $V_{15}$ – DODA were measured at 4.2 K in the sample with concentration $c = 4c_0$ by means of pulsed EPR Bruker E-580 X-band spectrometer operating at microwave frequency $\omega/2\pi$ = 9.7 GHz. Calibration of the microwave field amplitude $B_1$ was made using a standard isotropic $S$ = 1/2 radical. Symmetry based selection rules and evaluations of the intensities imply that transitions within the lowest four Zeeman sublevels of the doublet states have much smaller probabilities than those of the quartet states corresponding to the effective spin $S$ = 3/2 [1]. Rabi oscillations between $S$ = 3/2 states were induced by a long microwave field pulse of length *t* (Rabi pulse) and subsequently recorded by spin-echo detection giving access to the time evolution of the sample magnetization $\langle M_z(t) \rangle$. In both *t* pulse and $\pi/2 - \pi$ Hahn spin echo sequence the same $B_1$ values were used. Figures 1 and 2 show $\langle M_z(t) \rangle$ dependence for $B_0$=0.354 T and $B_1$ in the range 0.054 ÷ 1.24 mT (or, equivalently, for Rabi frequencies $\Omega_R/2\pi$ = 2.6 ÷ 59.2 MHz, where the ratio $\Omega_R/2\pi B_1$ = 48 MHz/mT was determined experimentally). Each $\langle M_z(t) \rangle$ corresponds to the superposition of oscillations obtained from all three transitions between $S$=3/2 states with slightly different Rabi frequencies (see [12]). All curves show a fast decrease at short times due to dephasing of spin packets with different resonance frequencies in an inhomogeneously broadened EPR line, followed by a number $n_R$ of damped Rabi oscillations. The damping time $\tau_R$ is very sensitive to $\Omega_R$, particularly in the range $8 \div 15$ MHz where the fastest decay ($n_R < 3$) is observed. In order to extract the damping time $\tau_R$ associated with each $\Omega_R$, each measured oscillating curve in Fig. 1 is fitted to $j_0(\Omega_R t)e^{-t/\tau_R}$, where $j_0(z) = \int_z^\infty J_0(z)dz$ ($J_0$ is the zero order Bessel function) is associated with the distribution of Larmor frequencies within the EPR line [5], while the damping of oscillations related to decoherence processes is given by a simple exponential $e^{-t/\tau_R}$. The full evolution of the damping rate $\tau_R^{-1}$ vs $\Omega_R$ is shown as a set of symbols in Fig. 3. The broad peak appearing in the range 8 MHz $< \Omega_R/2\pi <$ 15 MHz implies the existence of a new and efficient decoherence mechanism that is extremely sensitive to the microwave field amplitude. The amplitude of oscillations in that particular region does not obey a simple exponential law which results in large error bars in Fig. 3. The peak value $\tau_R = 36$ ns obtained at $\Omega_R/2\pi = 8$ MHz is nearly by an order of magnitude shorter than $\tau_2 = 250$ ns measured at the same experimental conditions. The slow linear variation of $\tau_R^{-1}$ with $\Omega_R/2\pi$ in the range 20-60 MHz is a consequence of the random distribution of the Landé factor of $V_{15}$ clusters in a frozen solution and intercluster dipolar interactions that result in a total decay rate $\sim 0.02\Omega_R$ (see [12]).

More detailed inspection shows that the decoherence peak around 8 MHz has a shoulder at $\Omega_R/2\pi \sim 15$ MHz, close to the Larmor frequency of protons of DODA in the resonant magnetic field $B_0$ ($\omega_N/2\pi = 15.1$ MHz for $B_0=0.354$ T). This suggests a decoherence mechanism associated with resonant electron-nuclear cross-relaxation when the electronic Rabi frequency $\Omega_R$ is close to the average proton Larmor frequency $\omega_N$. Such a mechanism of polarization transfer from the electronic to the nuclear spin bath is analogous to the one which takes place in Nuclear Spin Orientation Via Electron spin Locking (NOVEL) technique of dynamic nuclear polarization that is produced under resonant microwave field [16, 17] and with the electronic spin nutation frequency tuned to $\omega_N$ (Hartmann-Hahn condition [18]). However, here we are not interested in a high degree of polarization of the nuclear spin bath requiring a special sequence of pulses, but in the relatively low degrees of depolarization of the electronic spin bath after a single Rabi pulse. Such a depolarization leads to the strong decoherence mechanism in $V_{15}$ SMM which is seen and investigated here for the first time.

Because of the zero-field splitting $\Delta_{3/2} > \Omega_R$, at a given field, only one of three possible $S=3/2$ transitions in each $V_{15}$ cluster is actually induced by the microwave field, and we can use an effective spin-1/2 approximation. The Hamiltonian of this "central spin" [19] interacting with static and microwave external magnetic fields and a large number of nuclear spins can be written as follows (in units of Planck constant):

$$H = \omega_e S_z + 2\Omega_R S_x \cos\omega t + \sum_j \omega_j I_z^j + \sum_{\substack{j \\ \alpha=x,y,z}} A_{z\alpha}^j S_z I_\alpha^j, \quad (1)$$

where $\omega_e$ and $\Omega_R$ are correspondingly the Larmor precession frequency and the Rabi frequency of the central spin $S = 1/2$, $\omega_j$ are the precession frequencies of the nuclear spins $I = 1/2$ (in our case protons) distributed around $\omega_N$ with half-width $\sigma_N$. The $\sigma_N$ value may be regarded as the average local field (in frequency units) produced by the central spin at the site of the neighboring nuclear spin. The last term in Eq. (1) represents the super-hyperfine interaction between the central spin and nuclear spins where we have omitted the non-diagonal terms $\sim S_x, S_y$. After subsequent transformation of Eq. (1) to the rotating reference frame (see [12]), we obtain the following effective Hamiltonian:

$$H' = \Omega \tilde{S}_x + V(t); \quad V(t) = \frac{1}{2\Omega}\left(\Omega_R \tilde{S}_z + \varepsilon \tilde{S}_x\right) \sum_j \left\{ \left[\left(A_{zx}^j - i A_{zy}^j\right) e^{i\omega_j t} I_+^j + c.c.\right] + 2 A_{zz}^j I_z^j \right\}, \quad (2)$$

where $\Omega = \sqrt{\varepsilon^2 + \Omega_R^2}$ is the nutation frequency of the central spin ($\varepsilon = \omega_e - \omega$ is the shift of its precession frequency from $\omega$), and the following notations of the spin operators are used: $I_\pm^j = I_x^j \pm i I_y^j$, $\tilde{S}_x = (\Omega_R S_x + \varepsilon S_z)/\Omega$, $\tilde{S}_y = S_y$, $\tilde{S}_z = (\Omega_R S_z - \varepsilon S_x)/\Omega$. The term $\Omega \tilde{S}_x$ is the residual interaction of the central spin with the external magnetic field in the rotating reference frame. The perturbation $V(t) = \tilde{\omega}_x(t) \tilde{S}_x + \tilde{\omega}_z(t) \tilde{S}_z$ involves random local fields induced by the nuclei at the site of the central spin. These local fields have two components:

(i) $\tilde{\omega}_x(t)$ which is aligned in the direction of the central spin quantization axis $\tilde{x}$. The corresponding terms $\tilde{S}_x I_{+(-)}^j$ and $\tilde{S}_x I_z^j$ of the Hamiltonian (2) result mainly in spin dephasing. They are relevant only for spins far from resonance (with $\varepsilon \sim \Omega_R$);

(ii) transverse component $\tilde{\omega}_z(t)$ that comes from $\tilde{S}_z I_z^j$ and $\tilde{S}_z I_{+(-)}^j$ terms in the Hamiltonian (2). The term $\tilde{S}_z I_z^j$ is responsible for the second-order process of dephasing, and the corresponding damping rate of Rabi oscillations is limited by the rate of the nuclear bath internal dynamics [20]. The cross-relaxation terms $\tilde{S}_z I_{+(-)}^j$ are responsible for the mutual flip of the electronic and nuclear spins with respect to their quantization axes, leading to energy dissipation. These resonant processes occur only when $V(t)$ has spectral components close to the central spin nutation frequency, i.e. for $\omega_j \sim \Omega$. As opposed to spin dephasing, electron-nuclear cross-relaxation takes place even for slow nuclear spin baths, provided that the condition $\omega_j \sim \Omega$ is fulfilled.

In order to simulate the recorded Rabi oscillations, we have considered an ensemble of central spins with different $\varepsilon$ that comprise the inhomogeneously broadened EPR line and have obtained the following expression for the probability of a transition of the central spin determined by cross-relaxation processes [12]

$$P(\Omega,t) = \frac{\sigma_e^2 \Omega_R^2}{\Omega^2} \int d\omega \rho(\omega) \left\{ \frac{\sin^2[(\Omega-\omega)t/2]}{(\Omega-\omega)^2} + \frac{\sin^2[(\Omega+\omega)t/2]}{(\Omega+\omega)^2} \right\} \quad (3)$$

(here $\rho(\omega)$ is the distribution density of nuclear frequencies) that depends on the average local field $\sigma_e$ produced by the neighboring nuclear spins at the site of the central spin. At the moment $t$ the fraction of spins in the spin packet that are still coherently driven by the resonant microwave pulse is $e^{-P(\Omega,t)}$. Averaging over this ensemble of spin-packets leads to the following evolution of the recorded z-projection of magnetization:

$$M_z(t) \sim \int d\varepsilon g(\varepsilon) \frac{\varepsilon^2 + \Omega_R^2 \cos\Omega t}{\Omega^2} e^{-P(\Omega,t)}, \quad (4)$$

where $g(\varepsilon)$ is the spectral density of electronic spin states. At short times $P(\Omega,t) \sim 0$, and the time-dependent term in expression (4) is roughly proportional to $(\Omega_R t)^{-1/2}$ explaining the fast initial decrease of $\langle M_z(t) \rangle$ in terms of dephasing of different spin packets (see also the caption to Fig. 2). At longer times the damping of oscillations comes from the cross-relaxation factor $e^{-P(\Omega,t)}$. Equation (4) was used for numerical simulations of oscillation decays and Rabi times associated with each $B_1$ (thick red curves in Figs. 2 and 3). Correct estimation of the model parameters $\sigma_e$, $\sigma_N$ requires precise calculation of the local fields produced by the dipolar interactions at the vanadium and proton sites. Taking into account magnetic dipolar interactions between the 15 $V^{IV}$ ions within each $V_{15}$-DODA cluster and the nearest neighbor protons, we have obtained the following rough estimates: $\sigma_e \approx 2\pi \cdot 8$ MHz and $\sigma_N \approx 2\pi \cdot 2$ MHz. The damping factor $e^{-(\beta\Omega_R + \Gamma_2)t}$ (the dashed line in Fig. 3) was added in our calculations in front of the cosine in Eq. (4) in order to account for (i) the microwave-independent decoherence ($\Gamma_2 = 4.5$ µs$^{-1}$ attributed to $^{51}$V nuclear spin bath [1]), (ii) the dispersion of the *g*-factors in the disordered sample, and (iii) inter-cluster dipolar interactions. Both (ii) and (iii) result in the linear dependence of $\tau_R^{-1}$ upon $\Omega_R$ with the coefficient $\beta \approx 0.02$ (see [12]). Note that the $\Gamma_2$ value corresponds to coherence time $1/\Gamma_2 = 220$ ns which is close to the phase coherence time $\tau_2 = 250$ ns independently measured at the same conditions. Note that molecular vibrations have no such resonant effect as magnetic dipolar couplings and become important only at temperatures much higher than 4 K, i.e. when the corresponding spin-lattice relaxation time $\tau_1$ is comparable with the Rabi time $\tau_R$.

As seen in Fig. 3, the distinct features of the observed $\tau_R^{-1}(\Omega_R)$ dependence (a peak at 8 MHz and a shoulder at 15 MHz) are reproduced in calculations within the framework of the above model. Since for a given spin packet with $\varepsilon \neq 0$ its nutation frequency $\Omega > \Omega_R$, the Hartmann-Hahn condition $\Omega = \omega_N$ yields the condition $\Omega_R < \omega_N$. The direct analysis of $P(\Omega,t)$ dependence (3) gives us the position of the peak at $\Omega_R \sim \omega_N - \sigma_e$, explaining its shift from 15.1 MHz to 8 MHz (see [12]). The shoulder at $\Omega_R/2\pi = 15$ MHz comes from those V$_{15}$ spins that are close to resonance ($\Omega = \Omega_R$) and are responsible for Rabi oscillations at $\Omega_R t \gg 1$ when less coherent spin packets ($\Omega > \Omega_R$) are dephased sufficiently.

In conclusion, this first study of the decoherence of Rabi oscillations *vs* variable microwave field $B_1$ in a molecular magnet (V$_{15}$ embedded in the surfactant DODA) shows two distinct types of behavior in both our experimental and theoretical approaches: (i) a linear increase of Rabi decay rate $\tau_R^{-1}$ with $\Omega_R/2\pi > 15$ MHz associated with the dispersion of the *g*-factors of the V$_{15}$ clusters and with intercluster dipolar interactions, and (ii) a broad peak in the range $8 < \Omega_R/2\pi < 15$ MHz with a maximum near 8 MHz and a shoulder near 15 MHz. While (i) is rather well understood, (ii) evidences in favor of a new decoherence mechanism in the presence of microwave field. It generates polarization transfer between the electronic and nuclear subsystems accompanied with energy dissipation from electronic to nuclear spin baths. This mechanism is very general since it is observed for rather wide range of $\Omega_R$ values accounting for the inhomogeneous distribution of V$_{15}$ Larmor frequencies. The contribution to this decoherence mechanism from other nuclear spins ($^{75}$As, $^{14}$N and $^{51}$V) is negligible for $\Omega_R > 5$ MHz but, nevertheless, can play an important role when $\Omega_R \to 0$, i.e. in the absence of microwave pulse, as in $\tau_2$ measurements [1]. The decoherence with dissipation window revealed in this work should also be relevant for other qubit systems. For example, it should be pronounced in other proton-abundant SMMs like $Fe_8$ and $Mn_{12}$. Each $Fe_8$ cluster contains more than 100 protons, for which the existing experimental data on proton-induced super-hyperfine fields [21] and proton NMR [22] suggest $\sigma_e/2\pi \sim 7$ MHz and $\sigma_N/2\pi \sim 2$ MHz, i.e. a broad dissipation window of ~ 7 MHz. In the SMM $Mn_{12}$ $\sigma_N/2\pi \sim 2$ MHz [23], and rough estimates provide us with $\sigma_e/2\pi \sim 30$ MHz, so a very broad decoherence peak is expected here. A proper choice of $B_0$ and $B_1$ field values, evaluated from the present model, would bring $\Omega_R$ outside the window $[\omega_N - \sigma_e, \omega_N]$. That would elongate considerably the coherence times of the above SMMs in the presence of microwaves and would avoid a painstaking and in many cases impossible procedure of deuteration.

Authors acknowledge the CEA nano-science program for a one year post-doctoral grant, Kazan Federal University (project F11-22), Dynasty Foundation, and thank Noélie Marcellin (Alain Ibanez group, Néel Institute) for helping with processing samples in an oxygen-free glove box. S.B. thanks City of Marseille for financial support. B.T. thanks the Israel Science Foundation for financial support (ISF grant N° 168/09). A.M. thanks the Deutsche Forschungsgemeinschaft for continuous support.

\* Present address: Fakultät Physik, Technische Universität Dortmund, Dortmund, Germany
§ Present address: University of Liverpool, Dept. of Chemistry, Crown Street, L69 7ZD, UK.

**References**

[1] S. Bertaina et al., Nature **453**, 203 (2008). Corrigendum: Nature **466,** 1006 (2010).

[2] S.Takahashi et al., Nature **476**, 76 (2011).

[3] I. I. Rabi, Phys. Rev. **51**, 652 (1937).

[4] R. Boscaino et al., Phys. Rev. B **48**, 7077 (1993). R. N. Shakhmuratov et al., Phys. Rev. Lett. **79**, 2963 (1997).

[5] E. I. Baibekov, JETP Lett. **93**, 292 (2011).

[6] S. Bertaina et al, Nature Nanotechnology **2**, 39 (2007).

[7] M. A. Nielsen and I. L. Chuang, Quantum computation and quantum information (Cambridge University Press, Cambridge, 2000), P. 280.

[8] N. Prokof'ev and P. Stamp, J. Low Temp. Phys. **104**, 143 (1996).

[9] C. Schlegel et al., Phys. Rev. Lett. **101**, 147203 (2008).

[10] A. Ardavan et al., Phys. Rev. Lett. **98**, 057201 (2007).

[11] A. Tarantul, B. Tsukerblat, A. Müller, Inorganic Chemistry **46,** 161-169 (2007). A. Tarantul and B. Tsukerblat, Inorganic Chimica Acta **363**, 4361-4367 (2010). B. S. Tsukerblat and A. Tarantul, The nanoscopic $V_{15}$ cluster: an unique magnetic polyoxometalate, in "Molecular Cluster Magnets", ed. by R. E. P. Winpenny (World Scientific Publishers, Singapore, in press), Chap. 3, pp. 106-180.

[12] see supplementary material at http://link.aps.org/supplemental/

[13] D. Gatteschi et al., Nature **354**, 463 (1991).

[14] B. Barbara, J. Mol. Struct. **656,** 135 (2003).

[15] We have assumed a complete transfer of $V_{15}$ from the aqueous phase to the $CHCl_3$ phase. The assumption has been made based on complete de-coloration of the aqueous phase.

[16] A. Henstra et al., J. Magn. Res. **77**, 389 (1988).

[17] A. Henstra and W. Th. Wenckebach, Mol. Phys. **106**, 859 (2008).

[18] C. R. Hartmann and E. L. Hahn, Phys. Rev. **128**, 2042 (1962).

[19] N. Prokof'ev and P. Stamp, J. Low Temp. Phys. **104**, 143 (1996).

[20] V. V. Dobrovitski et al., Phys. Rev. Lett. **102**, 237601 (2009).

[21] W. Wernsdorfer et al., Phys. Rev. Lett. **84**, 2965 (2000).

[22] Y. Furukawa et al., Phys. Rev. B **64**, 094439 (2001).

[23] F. Borsa et al., Inorg. Chimica Acta **361**, 3777 (2008).

Figures

**FIG. 1**. Measured $\langle M_z(t) \rangle$ showing the evolution of Rabi oscillations as function of microwave field $B_1$ in the range $0.054 \div 1.24$ mT (or, equivalently, for Rabi frequencies $\Omega_R/2\pi = 2.6 \div 59.2$ MHz). As seen from the graphs, a strong damping of oscillations takes place in the region $8 \leq \Omega_R/2\pi \leq 15$ MHz.

**FIG. 2**. (Color online) Rabi oscillations in $V_{15}$–DODA recorded at Rabi frequencies $\Omega_R/2\pi = 30$ MHz (a) and $\Omega_R/2\pi = 10$ MHz (b). Experimental values and results of computations are represented by circles and thick lines, respectively. As the inhomogeneous EPR line half-width of this system $\sigma \sim 2\pi \cdot 700$ MHz by far exceeds $\Omega_R$, the mean $z$-component of magnetization is expected to follow the law $\langle M_z(t) \rangle \sim j_0(\Omega_R t) e^{-t/\tau_R}$, where $j_0(z \gg 1) \approx (2/\pi z)^{1/2} \cos(z + \pi/4)$ [5]. This form of $\langle M_z(t) \rangle$ was used in order to extract the experimental values of $\tau_R$ (110 ns and 48 ns for the curves (a) and (b), respectively). The theoretical curves computed according to Eq. (4) show a phase shift $\sim \pi/4$ relative to experimental points during the first period of oscillations which is due to the selective nature of spin-echo registration. Only a part of spin packets close to resonance contributes to the measured signal, thus reducing the half-width $\sigma$. However, this phase discrepancy does not affect the signal at longer scale ($\Omega_R t > 2\pi$), as well as Rabi times $\tau_R$.

**FIG. 3**. (Color online) The decay rate of Rabi oscillations $1/\tau_R$ vs Rabi frequency $\Omega_R/2\pi$. Measured and calculated data are represented as squares and the solid line, respectively. The dashed line represents the linear dependence $0.02\Omega_R + 4.5 \cdot 10^6$ s$^{-1}$ (see text).

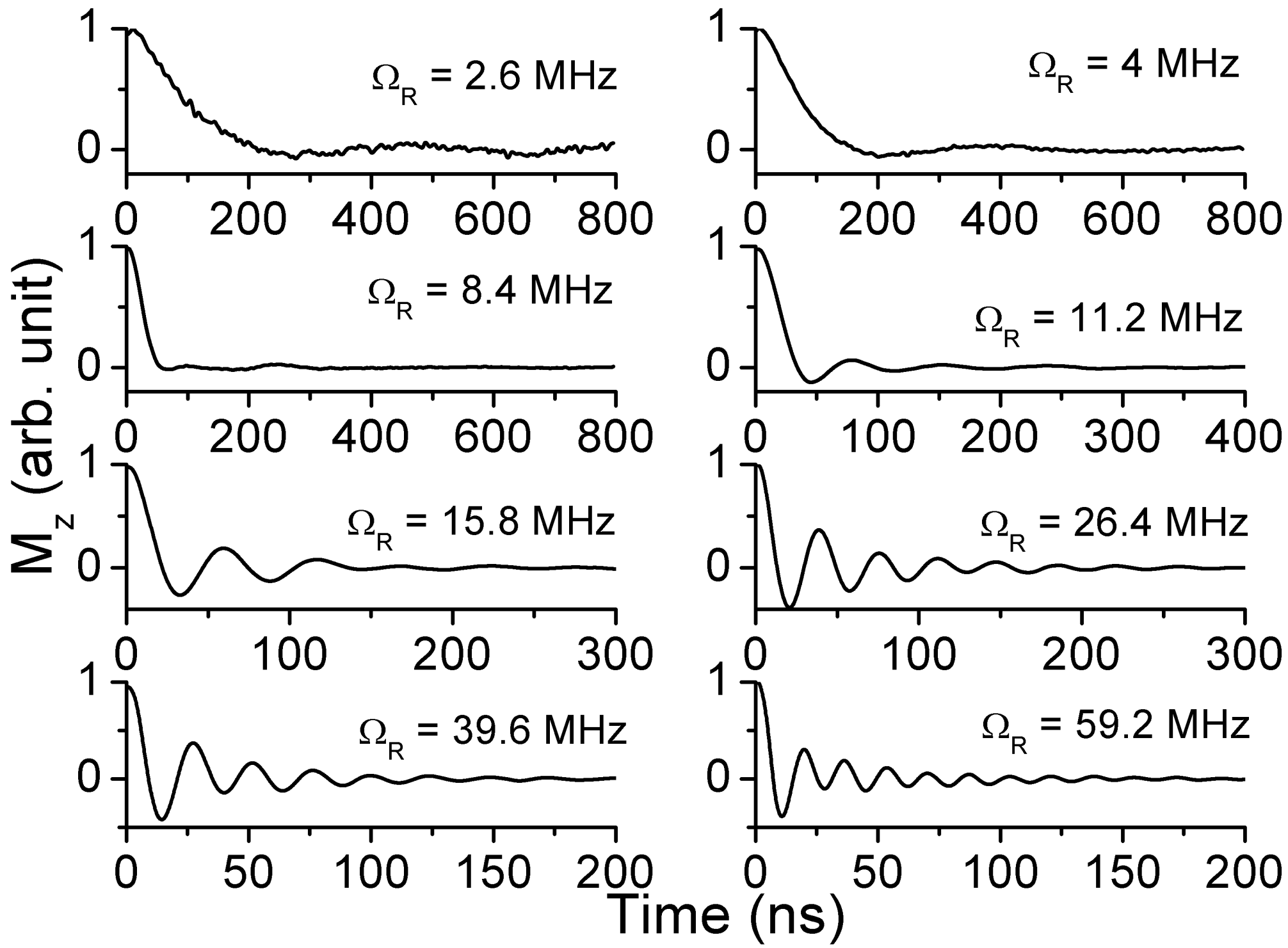

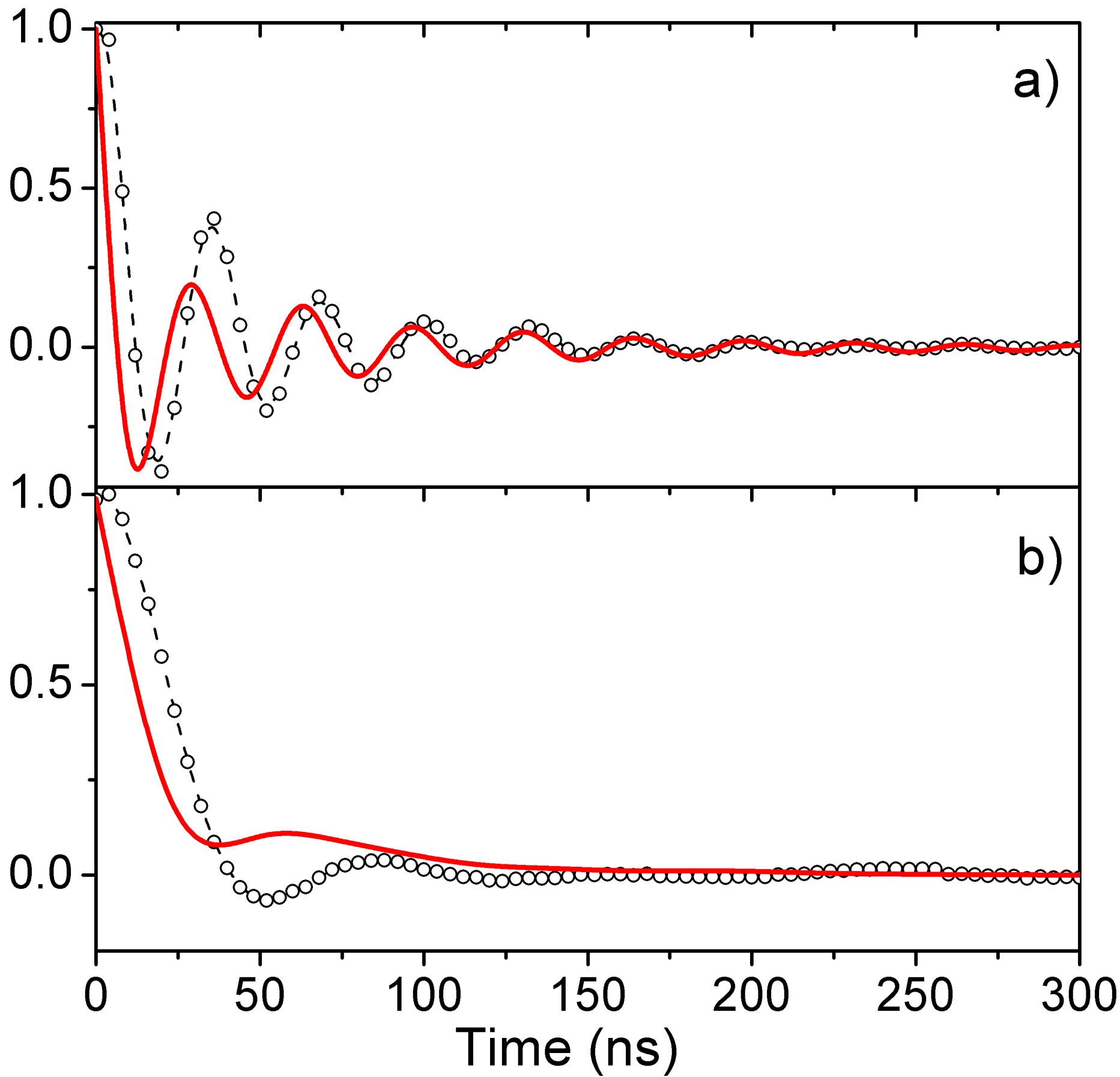

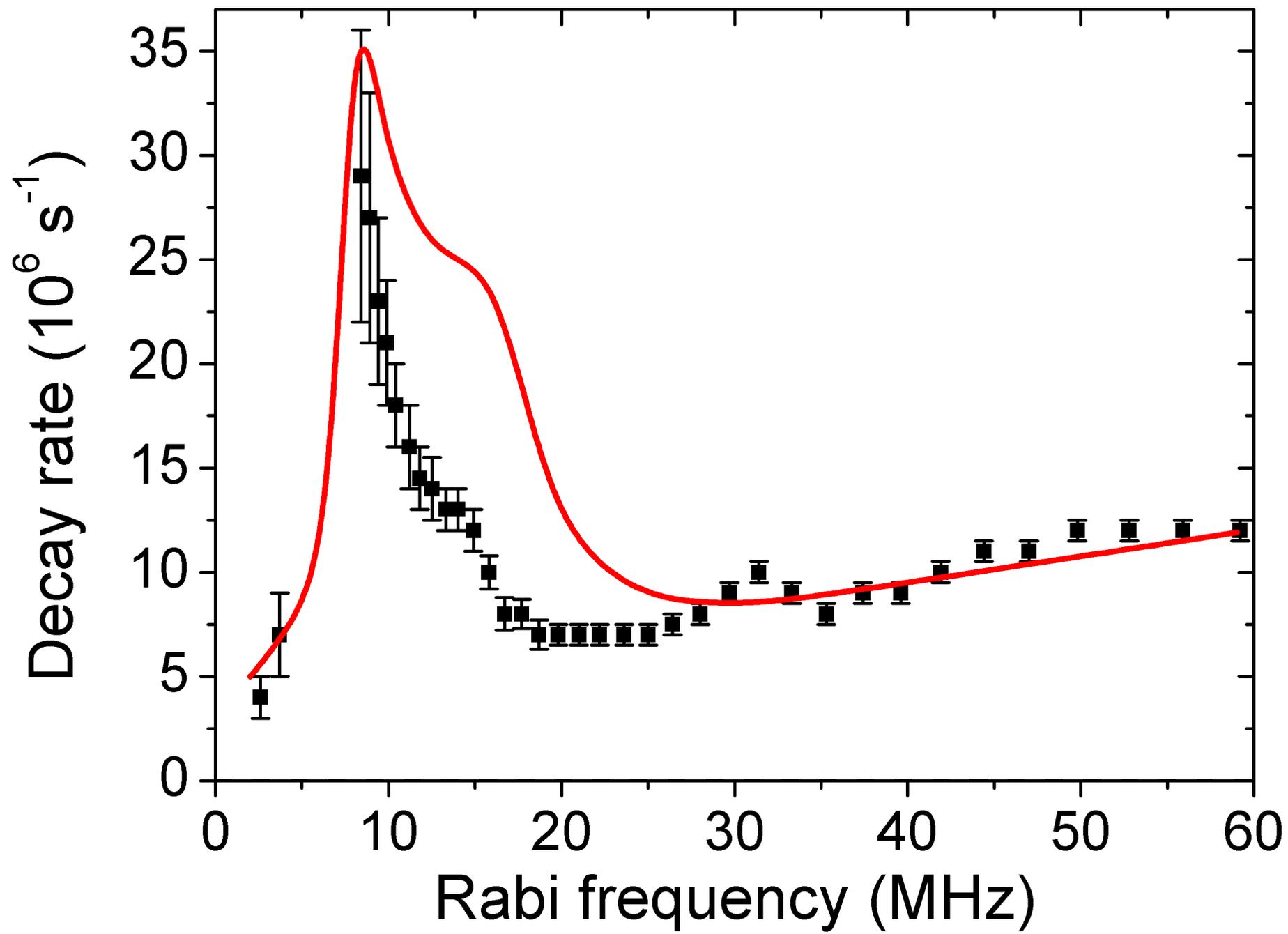